\documentclass[a4paper,fleqn]{cas-dc-arxiv}

\usepackage[numbers]{natbib}
\usepackage{lineno,hyperref}

\usepackage{epsfig}

\usepackage{amssymb}
\usepackage{amsthm}

\usepackage{subfigure}
\usepackage{graphics}
\usepackage{multicol}
\usepackage{graphicx}

\modulolinenumbers[5]

\def\tsc#1{\csdef{#1}{\textsc{\lowercase{#1}}\xspace}}
\tsc{WGM}
\tsc{QE}
\tsc{EP}
\tsc{PMS}
\tsc{BEC}
\tsc{DE}

\begin{document}
\let\WriteBookmarks\relax
\def\floatpagepagefraction{1}
\def\textpagefraction{.001}
\shorttitle{Suspensions of magnetic nanogels at zero field}
\shortauthors{I. S. Novikau et~al.}

\title [mode = title]{Suspensions of magnetic nanogels at zero field: equilibrium structural properties}                      



\author[1]{Ivan S. Novikau}
\cormark[1]
\ead{ivan.novikau@univie.ac.at}

\credit{TBD} 

\address[1]{Computational Physics, University of Vienna, Sensengasse 8, Vienna, Austria.}

\author[1]{Elena S. Minina}

\credit{TBD} 

\author[2,3]{Pedro A. S\'anchez}[orcid=0000-0003-0841-6820]
\ead{r.p.sanchez@urfu.ru}

\credit{TBD} 

\address[2]{Ural Federal University, 51 Lenin av., Ekaterinburg, 620000, Russian Federation.}
\address[3]{Institute of Ion Beam Physics and Materials Research, Helmholtz-Zentrum Dresden-Rossendorf e.V., D-01314 Dresden, Germany.}

\author[1,2]{Sofia S. Kantorovich}[orcid=0000-0001-5700-7009]
\ead{sofia.kantorovich@univie.ac.at}

\credit{TBD} 

\cortext[cor1]{Corresponding author}

\begin{abstract}
Magnetic nanogels represent a cutting edge of magnetic soft matter research due to their numerous potential applications. Here, using Langevin dynamics simulations, we analyse the influence of magnetic nanogel concentration and embedded magnetic particle interactions on the self-assembly of magnetic nanogels at zero field. For this, we calculated radial distribution functions and structure factors for nanogels and magnetic particles within them. We found that, in comparison to suspensions of free magnetic nanoparticles, where the self-assembly is already observed if the interparticle interaction strength exceeds the thermal fluctuations by approximately a factor of three, self-assembly of magnetic nanogels only takes place by increasing such ratio above six. This magnetic nanogel self-assembly is realised by means of favourable close contacts between magnetic nanoparticles from different nanogels. It turns out that for high values of interparticle interactions, corresponding to the formation of internal rings in isolated nanogels, in their suspensions larger magnetic particle clusters with lower elastic penalty can be formed by involving different nanogels.  Finally, we show that when the self-assembly of these nanogels takes place, it has a drastic effect on the structural properties even if the volume fraction of magnetic nanoparticles is low.
\end{abstract}


\maketitle

\sloppy

\section{Introduction}
The concept of microgel is almost 70 years old \cite{Baker1949}. The term refers to a colloidal soft particle made of a permanently crosslinked network of polymers, whose size can range from tens of nanometers to several micrometers \cite{2011-fernandez-nieves-bk, 2017-hamzah-jpr}. Microgels can be made responsive to different stimuli, as temperature, pH or external fields \cite{Gorelikov2004,Mohanty2015,Kobayashi2016,Backes2017a}. As a result, they have a great potential for many technological and bio-medical applications \cite{Son2016,PU2019}. Microgels under $\sim$100~nm in diameter, also known as nanogels, are especially promising for drug delivery \cite{Vinogradov2010}. Nowadays, the amount of studies devoted to these systems is growing very fast \cite{2019-rovigatti-sm-rev, 2019-martin-molina-jml-rev}.

Among the different responsive behaviours obtained for micro- and nanogel particles, the response to external magnetic fields is particularly appealing. This is achieved experimentally by embedding magnetic nanoparticles (MNPs) into the polymer network \cite{2013-jiang,2015-backes-jpcb}. Despite such interest, the theoretical understanding of magnetic micro- and nanogels is still rather limited due to the challenges involved in their modeling.

Recently, we introduced a coarse-grained computer simulation model of magnetic nanogels that allowed us to study the influence of the magnetic filler concentration on the structure of single nanogel particles \cite{Minina2018}. Regarding the polymer network, the model represents qualitatively the internal structure of nanogels obtained by electrochemically or photonically induced crosslinking of polymer precursors confined in emulsion nanodroplets \cite{2016-galia-eccm,2016-mavila-chrv,2019-minina}.

In this study we employ the aforementioned coarse-grained nanogel model to investigate the equilibrium structural properties of magnetic nanogel suspensions in absence of an applied external field. To the best of our knowledge, no theoretical study on magnetic nanogel suspensions is yet available in literature.

The structure of the paper is the following. First, in Section \ref{sec:model}, we briefly describe the model and simulation method. In Section \ref{sec:rnd}, we discuss the influence of nanogel concentration and the impact of magnetic interaction strength on the structural properties of nanogels suspensions. Finally, a brief summary of the work can be found in Section \ref{sec:conc}.

\section{Simulation Approach}\label{sec:model}

Our nanogel model is based on a bead-spring representation of the polymer chains and the embedded magnetic particles \cite{2019-minina}. Briefly, the setup of each nanogel particle is performed in the following way. First, polymer precursors are modelled as chains of spherical beads with unit dimensionless mass and diameter. They have a steric repulsion given by a truncated and shifted Lennard-Jones potential, or Weeks-Chandler-Andersen (WCA) potential \cite{1971-weeks}:
\begin{equation}
U_{W C A}(r)=\left\{\begin{array}{ll}{4\left[r^{-12}-r^{-6}\right]+1,} & {r \leqslant 2^{1/6}} \\ {0,} & {r > 2^{1/6}}\end{array}\right. ,
\label{eq:wca}
\end{equation}
where $r$ is the centre-to-centre distance between the interacting particles. Here, we set the depth of the Lennard-Jones potential well to unity, thus introducing a scaling for all energies in the simulation protocol. These beads form the polymer chain backbones by means of FENE springs connected to their centres:
\begin{equation}
U_{F E N E}(r)=-\frac{1}{2} \epsilon_{f} r_f^2 \ln \left[1-\left(\frac{r}{r_{f}}\right)^{2}\right],
\label{eq:fene}
\end{equation}
where $\epsilon_{f}=22.5$ is the dimensionless interaction strength and $r_{f}=1.5$ is the maximum bond extension. Nanogel particles are obtained by equilibrating $N_{p}=6$ polymer chains with $L=100$ beads each inside a spherical confinement wall with volume fraction of approximately $\phi_p \approx 0.1$. After equilibration, interchain crosslinks are randomly introduced according to a minimum interparticle distance criterium, up to reach a fraction of crosslinks of $\phi_{\mathrm{links}}=0.17$. Each crosslink consists in a elastic spring connecting the centres of the newly bonded pair of particles. In order to speed up the crosslinking process, harmonic springs are used for this purpose:
\begin{equation}
U_{h}(r)=-\frac{1}{2} Kr^{2}.
\label{eq:harm}
\end{equation}
By using $K=10$, we ensured that the mechanical effect of these springs is equivalent to the FENE bonds connecting the precursor backbones. For further details on the crosslinking protocol see Reference~\cite{2019-minina}. Regarding the magnetic particles, for simplicity they are introduced by assigning a permanent magnetic dipole, $\vec \mu$, at the centre of randomly selected beads, up to a fraction of $\phi_{\mathrm{m}}=0.1$. Therefore, these magnetic beads interact by means of the dipole-dipole pair potential:
\begin{equation}
U_{d d}\left(\vec{r}_{i j}\right)=\frac{\left(\vec{\mu}_{i} \cdot \vec{\mu}_{j}\right)}{r^{3}}-\frac{3\left(\vec{\mu}_{i} \cdot \vec{r}_{i j}\right)\left(\vec{\mu}_{j} \cdot \vec{r}_{i j}\right)}{r^{5}},
\label{eq:dipdip}
\end{equation}
\begin{figure}
        \centering
        \includegraphics[width=0.29 \textwidth]{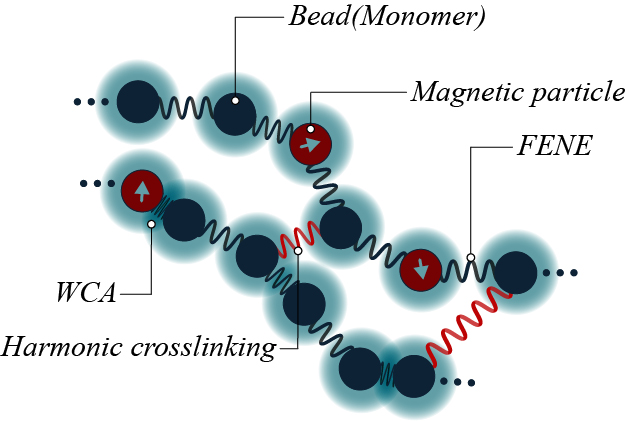}
        \caption{Bead-spring representation of the internal structure of a model magnetic nanogel. Arrows inside particles indicate the presence of a point magnetic dipole.}
        \label{fig:Model_beads}
\end{figure}
where $\vec \mu_i$, $\vec \mu_j$ are the respective dipole moments of the interacting particles and $\vec r_{ij}$ is the displacement vector connecting their centres.  These long range magnetic interactions were calculated using the dipolar P$^3$M algorithm \cite{2008-cerda-jcp}. Figure~\ref{fig:Model_beads} shows a sketch of this bead-spring representation of our model magnetic nanogel particles.

Finally, suspensions were simulated by placing 100 nanogel particles obtained from the procedure described above into a periodic cubic box with a volume fraction of beads fixed either to 0.1 or 0.2. It is worth noting that each individual nanogel was previously equilibrated. In total, 10 different equilibrium configurations with different cross-linker and magnetic particle intrinsic distributions were used to form the suspension. This allowed us to avoid the dependence of self-assembly on individual nanogel topology. Molecular dynamics simulations of such systems were performed with the simulation package {ESPREsSo} \cite{2019-weik}. A Langevin thermostat with fixed dimensionless temperature $T=1$ was used to mimic the thermal fluctuations of the background fluid. The system was first equilibrated by making $2 \cdot 10^{7}$ integration steps, using a fixed time step $\delta t=0.01$. subsequent measurements were obtained for $8 \times 10^{7}$ integration steps. Each set of parameters was sampled with 5 independent runs using different initial configurations.

Note that, in the system of dimensionless units defined above, the conventional dipolar coupling parameter, that measures the ratio between the dipole-dipole interactions and the thermal fluctuations, can be simply defined as $\lambda=\mu^2$.

\section{Results and Discussions}\label{sec:rnd}
\begin{figure}
        \centering
        \subfigure[]{\label{fig:snapshot}\includegraphics[width=0.25 \textwidth]{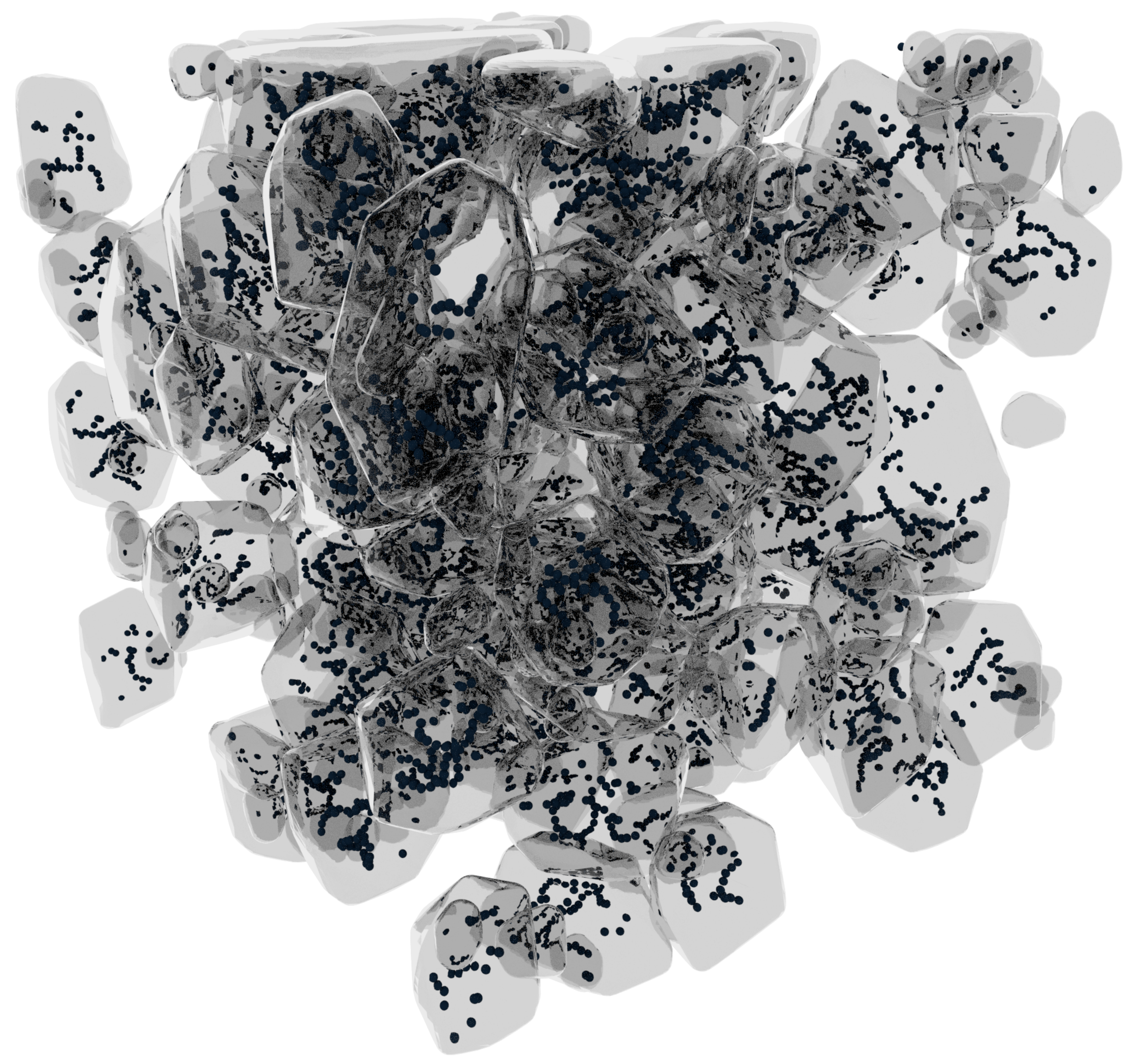}}
        \subfigure[]{\label{fig:self-assem}\includegraphics[angle = 90,width=0.12 \textwidth]{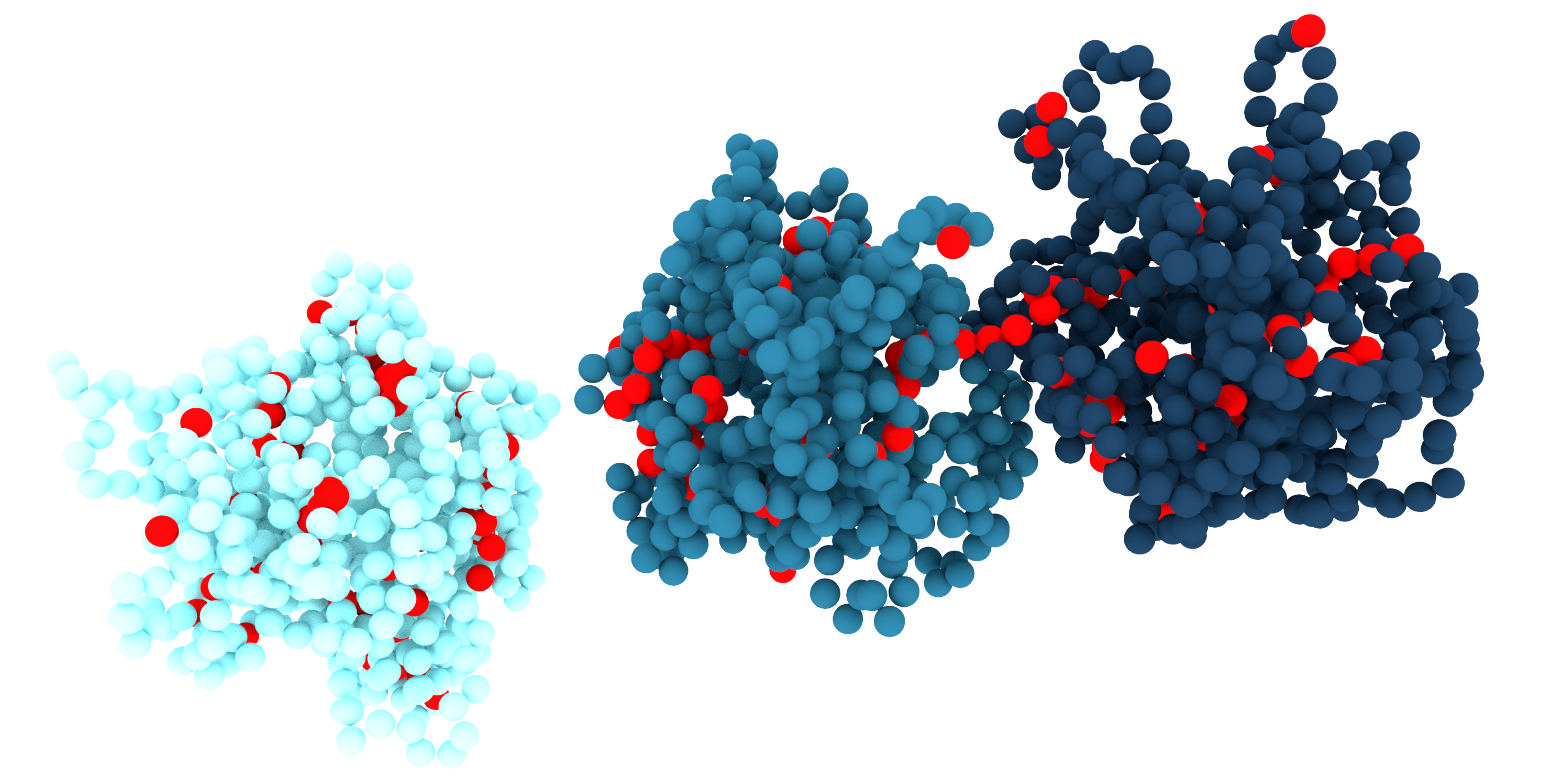}}
        \caption{(a) Simulation snapshot of a nanogel suspension. Volume fraction is 10 \%. $\lambda = 6$. Only magnetic particles are shown explicitly,  nanogels are represented by their convex hulls. (b) Zoomed in three nanogels with all beads shown explicitly. Magnetic particles are shown in red, nonnmagnetic beads of each nanogel have the same colour (light blue, blue and dark blue).}
        \label{fig:Model}
\end{figure}

We start the analysis with the visual inspection of the equlibrated suspensions. As long as the system is rather crowded, in Fig.~\ref{fig:snapshot}, we show only the magnetic particles explicitly, whereas the whole structures of each nanogel particle is represented by its convex hulls. This snapshot was obtained for $\lambda=6$ and overall volume fraction of 0.1. In the lowest part of the snapshot (close to the centre), one can find two nanogels that seem to form a cluster. From our previous work \cite{Minina2018}, it is known that inside an isolated  magnetic nanogel with $\lambda=6$, magnetic particles tend to self-assemble and form long chains close to the periphery of the nanogel to minimise the curvature. However, the translational motion required for self-assembly is often penalised by the elastic network. In case of suspension, dipolar energy can be additionally minimised if magnetic particles close to the surface of one nanogel form favourable contacts with magnetic nanoparticles of a neighbouring nanogel, thus, leading to the nanogel self-assembly, as shown in Fig.~\ref{fig:self-assem}.

\begin{figure}
	\centering
	\subfigure[]{\label{fig:rdf-cm-c}\includegraphics[width=0.23\textwidth]{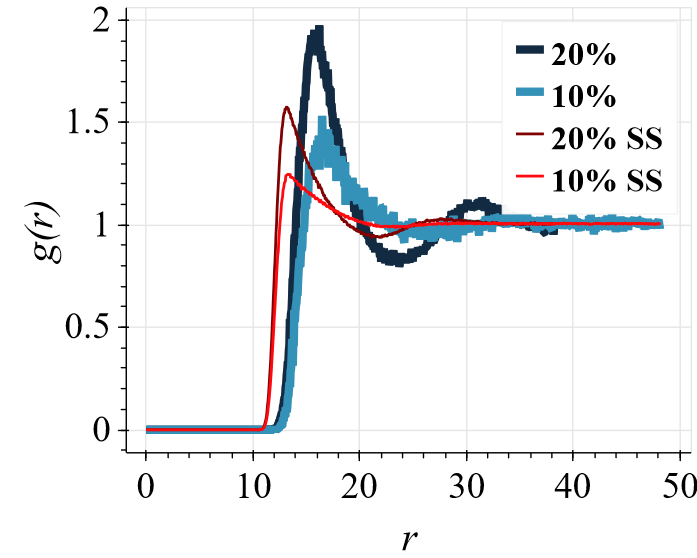}}
	\subfigure[]{\label{fig:rdf-cm-l}\includegraphics[width=0.23\textwidth]{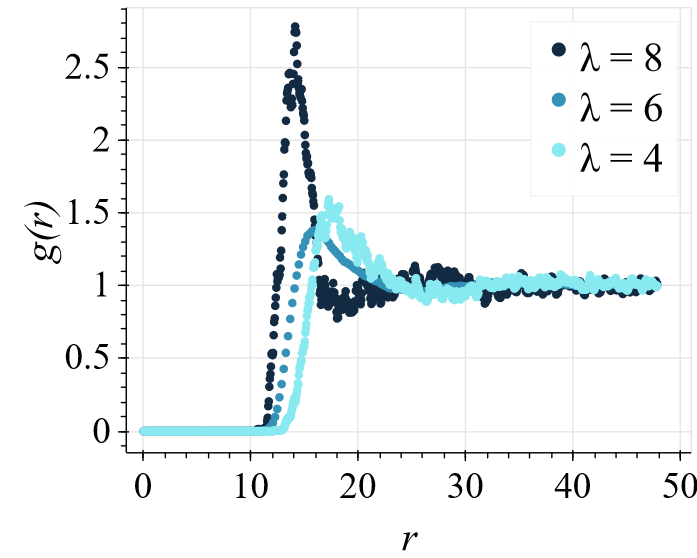}}
\subfigure[]{\label{fig:rdf-mag-c}\includegraphics[width=0.23\textwidth]{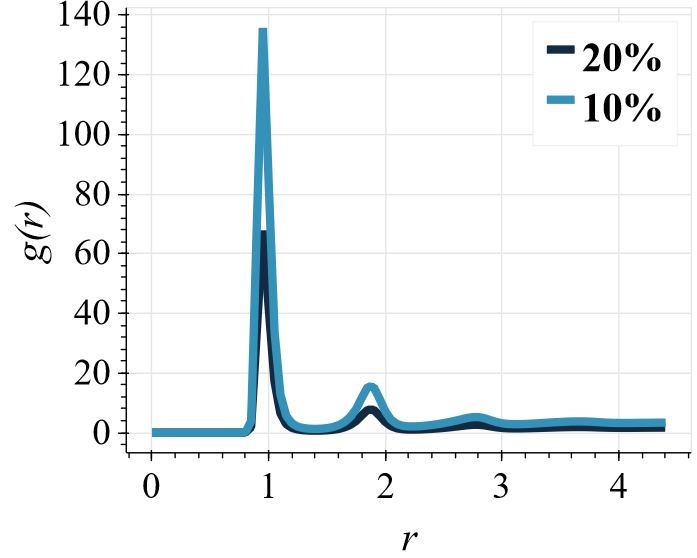}}
	\subfigure[]{\label{fig:rdf-mag-l}\includegraphics[width=0.23\textwidth]{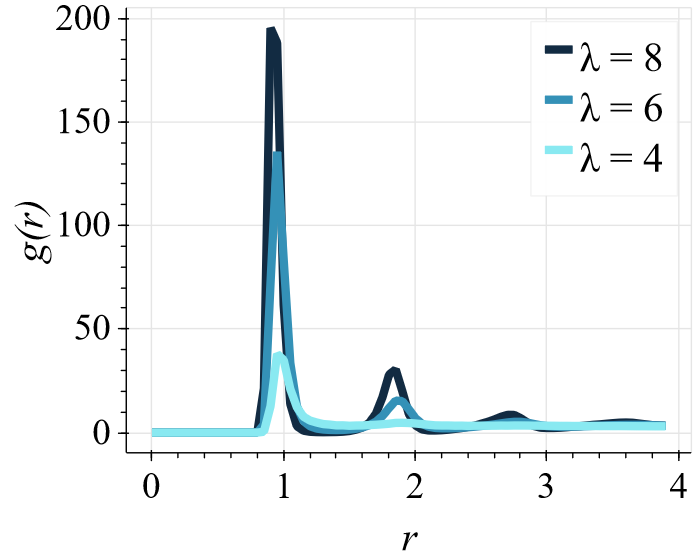}}
	\caption{Radial distribution functions (RDFs.) (a) and (b) -- RDFs calculated for the nanogels centres of mass. Nanogels radius of gyration is on average equal to six. (c) and (d) -- RDFs computed using exclusively magnetic particles. (a) and (c) show the differences in RDFs caused by changes in nanogel volume fraction; (b) and (d) show the impact of $\lambda$ on the RDFs. RDFs for equivalent suspensions of WCA spheres (SS) are also included in (a).}
	\label{fig:rdf-conc-l}
\end{figure}

In order to understand how intensively the nanogels self-assemble, in Figs.~\ref{fig:rdf-cm-c} and \ref{fig:rdf-cm-l}, we plot radial distribution functions calculated for nanogels centres of mass. For a volume fraction of 10\% (light blue curve, Fig.~\ref{fig:rdf-cm-c}), one can hardly see any signature of self-assembly. The first peak of the RDF is rather small and the first minimum is not very pronounced. The situation changes for higher nanogel concentrations. Doubling the volme fraction results in a very clearly pronounced first peak at $r=12$. It is worth mentioning here that the average nanogel radius of gyration is $R_{g}=6$ and remained unchanged during the simulation ($\pm5\%$ of the initial value). Thus, $r=12$ corresponds to the close contact of two nanogels similar to the one shown in Fig.~\ref{fig:self-assem}. To exclude the possibility of having simply density fluctuation effects captured by the RDF, we also calculate the RDFs of WCA spheres suspensions with radii equal to $R_g$ and plot them with thin red lines in Fig.~\ref{fig:rdf-cm-c}. The comparison between WCA spheres and actual magnetic nanogels clearly reveals the signature of self-assembly in the latter for the case 20\% volume fraction: the existence of a second maximum and a deep first minimum exhibited by the RDF of nanogel suspension.  Apart from concentration impact, self-assembly of magnetic nanogels can be also tuned by changing the interaction strength between MNPs. This is illustrated in Fig.~\ref{fig:rdf-cm-l}, where we plot the RDFs calculated for the centres of mass of nanogels with different interaction strengths for their magnetic beads. In case of $\lambda = 4$, the RDF has a perfect shape of a noninteracting WCA-sphere liquid. In contrast, for $\lambda = 8$ the self-assembly is clearly taking place. 

Further microscopic information can be revealed by calculating the RDFs corresponding only to the magnetic particles. This result is shown in Figs.~\ref{fig:rdf-mag-c} and \ref{fig:rdf-mag-l}. As expected, magnetic particles in nanogels and across them self-assemble and the tendency to form larger clusters grows with increasing nanogel concentration as well as with value of $\lambda$. Thus, for example, for $\lambda = 8$ (dark blue in Fig. \ref{fig:rdf-mag-l}), one finds well-defined peaks up to the fourth coordination shell, meaning that the chain-like magnetic structures formed in the system are rather frequent and have a significant length.

\begin{figure}
	\centering
	\subfigure[]{\label{fig:sf-cm-c}\includegraphics[width=0.23\textwidth]{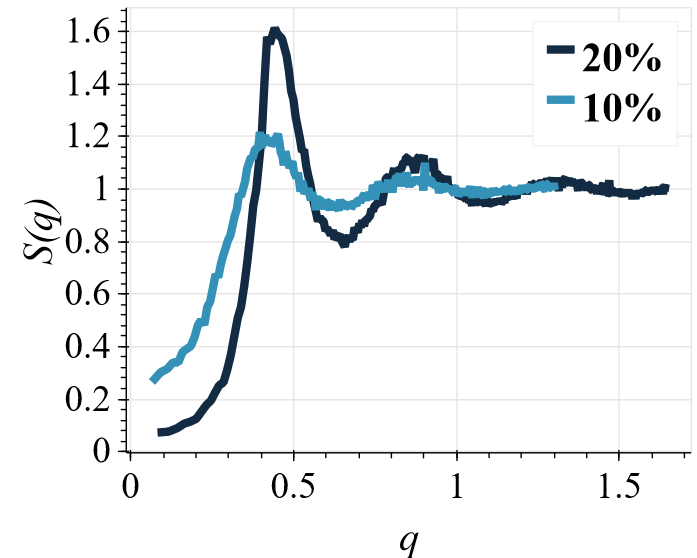}}
	\subfigure[]{\label{fig:sf-cm-l}\includegraphics[width=0.23\textwidth]{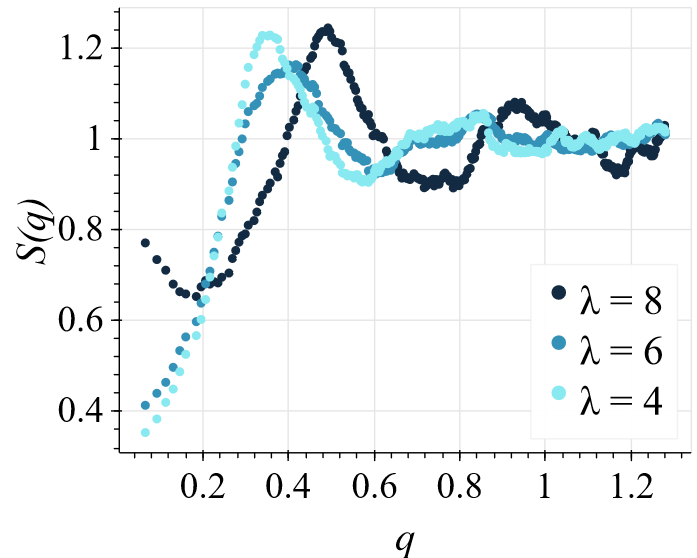}}
\subfigure[]{\label{fig:sf-mag-c}\includegraphics[width=0.23\textwidth]{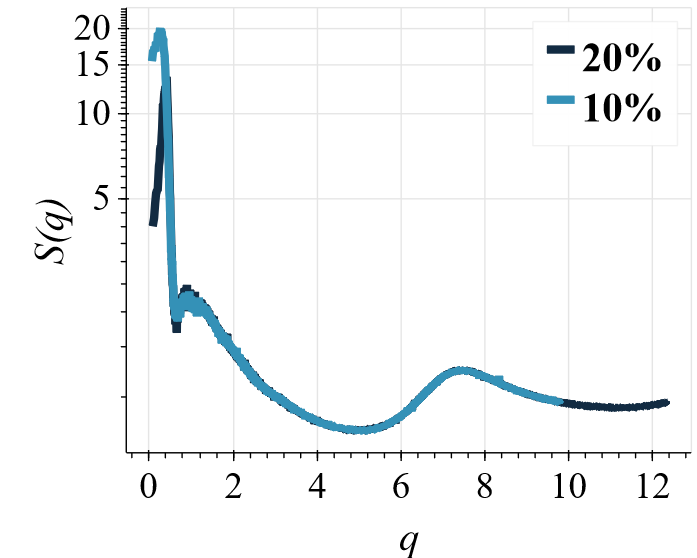}}
	\subfigure[]{\label{fig:sf-mag-l}\includegraphics[width=0.23\textwidth]{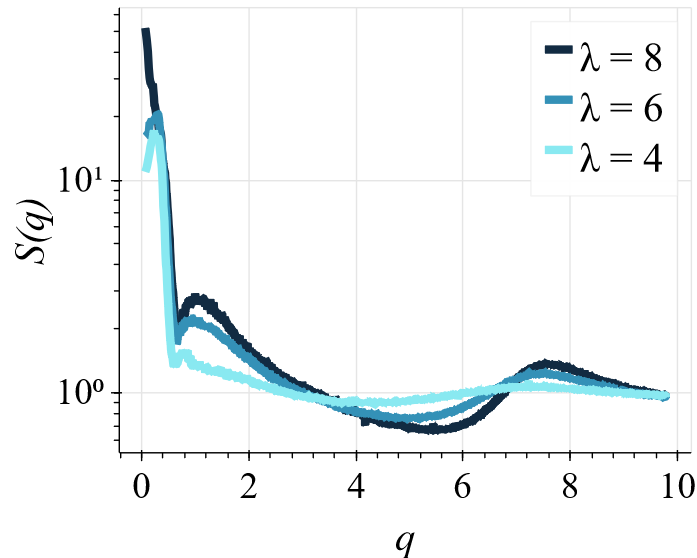}}
	\caption{Structure factors (SFs.) (a) and (b) -- SFs calculated for whole nanogels structures. (c) and (d) -- SFs computed using exclusively magnetic particles. (a) and (c) show the differences in SFs caused by changes in nanogel volume fraction; (b) and (d) show the impact of $\lambda$ on the SFs.}
	\label{fig:sf-conc-l}
\end{figure}

To answer the question whether the self-assembly of magnetic nanogels can be detected experimentally, we calculated their structure factors (SFs). Analogously to RDFs, we computed SFs for overall nanogels and for magnetic particles only. Note these SFs, presented in Fig.~\ref{fig:sf-conc-l}, are calculated directly from the simulation data using the procedure described in Reference~\cite{2009-pyanzina} and not as Fourier transform of the RDFs from Fig.~\ref{fig:rdf-conc-l}.
The first peaks, $q \sim 0.5$, in Figs.~\ref{fig:sf-cm-c} and \ref{fig:sf-cm-l} correspond to approximately $2R_g \sim 12$ in real space. They reflect the number of nanogel pairs in close contact. Their height grows and their position shifts slightly to the right with both, increasing concentration (\ref{fig:sf-cm-c}) and growing value of $\lambda$ (\ref{fig:sf-cm-l}). The shift of the peaks can be explained by two effects: first, the radius of gyration of magnetic nanogels decreases with $\lambda$ \cite{Minina2018}; second, when forming a contact through magnetic nanoparticles, these nanogels get deformed and can interpenetrate to a certain extent. The scale of these effects can be estimated from the shift of the SF first peak to be around 10\% for $\lambda =6 $ in case of growing volume fraction, or to be around 20 \% if the value of $\lambda$ changes from 6 to 8 and the volume fraction is fixed to 10\%. Fig.~\ref{fig:sf-mag-c} shows the SF of magnetic nanoparticles calculated for different volume fractions and fixed $\lambda = 6$. It is clearly seen that the overall volume fraction affects only the region of small $q$. In fact, doubling the volume fraction of magnetic nanogels seems to not affect qualitatively the  self-assembly of magnetic nanoparticles, as it was also seen in Fig.~\ref{fig:rdf-mag-c}. For $\lambda = 4$, the shape of SF (light blue curve, Fig.~\ref{fig:sf-cm-l}) suggests that the self-assembly is insignificant. This observation is confirmed by Fig.~\ref{fig:sf-mag-l}, where the peak at $q \sim 7$, corresponding to the close contact of two magnetic beads, is negligible. In contrast, for $\lambda=8$ we can clearly observe the latter peak. In Figs.~\ref{fig:sf-mag-c} and \ref{fig:sf-mag-l} SFs have 2 peaks for $q<2$. In real space this corresponds to  distances larger than $\sim 3$. Whereas the first peak at $q \sim 0.5$ is clearly related to the close contact of two nanogels at distance $2R_g \sim 12$, the second peak is more difficult to interpret. Corresponding to real distances on the order of $R_g$, most probably shows the correlation length of magnetic particles inside individual nanogels.

Finally, having analysed all evidences of magnetic nanogel self-assembly and described the qualitative trends, in Table \ref{table:mean-size} we collect mean clusters sizes measured for, both magnetic nanogels and individual magnetic particles, for all systems investigated here. For magnetic particles we used the distance-energy criteria of Reference~\cite{2009-pyanzina}, whereas for whole nanogels pairs we considered them to be connected only if they had a shared cluster of more than three magnetic particles, with at least two of them belonging to each of the nanogels. 
\begin{table}
\centering
\small
\begin{tabular}{c||ccc|c}  \hline\hline
     & $\lambda = 4$ & $\lambda = 6$, 10 \% &  $\lambda = 8$  & $\lambda = 6$, 20 \%  \\\hline
MNPs & 3.7 & 7.6 & 20.0 & 7.5\\
nanogels & 0.2 & 2.1 & 5.2 & 2.6\\ \hline
\end{tabular}
  \caption{Cluster sizes measured for MNPs only (upper row) and for whole nanogels (lower row). For cluster definition, see the main text. The errorbars for these values are below 5 per cent.}
  \label{table:mean-size}
\end{table}
One can see that all the implicit evidences described above are fully confirmed by the clusters sizes: we see no aggregation for magnetic nanogels for $\lambda =4$, even though, magnetic nanoparticles do agregate; for $\lambda = 6$ we see a moderate level of nanogel self-assembly enhanced by the increase of the volume fraction; for $\lambda = 8$ the nanogel self-assembly is very pronounced. It is worth noting here that a cluster of 5 nanogels is a relatively large object that cannot but affect rheological and mechanical properties of these suspensions. The study of these properties is currently in progress.  

\section{Conclusions}\label{sec:conc}

In this paper we thoroughly analysed the influence of magnetic nanogel concentration and magnetic particle interactions on the self-assembly of magnetic nanogels in zero field. Our results show that, whereas $\lambda \geq 4$ can be considered as strong self-assembly conditions for free magnetic nanoparticles, magnetic nanogels containing these particles only start forming clusters at $\lambda \geq 6$. It is worth reminding here that for individual magnetic nanogels, if the value of $\lambda$ exceeds 6-7, the initial susceptibility  decreases due to the fact that polymer matrix cannot hinder anymore the formation of rings of magnetic nanoparticles inside the nanogel \cite{Minina2018}. It was found here that this effect vanished in suspensions due to the possibility of forming larger magnetic clusters with lower elastic penalty involving particles from different nanogels in close contact. These connections were shown to be responsible for nanogel self-assembly. The values of $\lambda$ for which real nanogels will start forming bridges, containing magnetic particles, and self-assemble will, however, strongly depend on the crosslinking degree of the nanonogels, as the latter parameter was shown to have a nontrivial impact on the intrinsic magnetic particle self-assembly within individual nanogels \cite{Minina2018}. The formation of bridges, albeit only qualitatively, was reported recently in Ref. \cite{2019-witt} (see, Fig. 4a). The degree of clusterisation of magnetic nanoparticles inside individual nanogels, as well as between them, can be verified by scattering techniques \cite{2008-eckert,2011-galicia} as we show by calculating magnetic nanogel structure factors. Our findings revealed three length-scales inherent to magnetic nanoparticles in these systems. The largest scale (region of small wave vectors $q$) corresponds to twice the nanogel radius of gyration and can be attributed to correlations of magnetic nanoparticles from different nanogels forming clusters; the second length-scale corresponds to the nanogel radius of gyration and may reflect the correlation length of the magnetic nanoparticles inside one nanogel; finally the smallest scale, corresponding to high values of wave vectors $q$, shows close contacts of nanoparticles. Heights and positions of all three SF peaks were shown to be sensitive to the value of $\lambda$. Interestingly, only the region of small $q$ was found to depend on the nanogel volume fraction. The latter, however, plays a significant part in the amplitude of SFs if calculated for whole nanogel structures. Summarising our findings, one can expect cluster formation in suspension of magnetic nanogels, in absence of external magnetic fields, only if the interaction between embedded ferromagnetic nanoparticles is significantly higher than the values of thermal energy. However, if the self-assembly takes place, it might have a drastic effect on the structural properties even for relatively low volume fraction of the magnetic filler.

\section{Acknowledgements}
This research has been supported by the Russian Science Foundation Grant No.19-12-00209. Authors acknowledge support from the Austrian Research Fund (FWF), START-Projekt Y 627-N27. Computer simulations were performed at the Vienna Scientific Cluster (VSC-3).

\end{document}